# Vacuum acceleration of electrons in a dynamic laser pulse

D. Ramsey, P. Franke, T.T. Simpson, D.H. Froula, and J.P. Palastro
*University of Rochester, Laboratory for Laser Energetics, Rochester, New York 14623, USA*

**Abstract**

A planar laser pulse propagating in vacuum can exhibit an extremely large ponderomotive force. This force, however, cannot impart net energy to an electron: As the pulse overtakes the electron, the initial impulse from its rising edge is completely undone by an equal and opposite impulse from its trailing edge. Here we show that planar-like "flying focus" pulses can break this symmetry, imparting relativistic energies to electrons. The intensity peak of a flying focus—a moving focal point resulting from a chirped laser pulse focused by a chromatic lens—can travel at any subluminal velocity, forwards or backwards. As a result, an electron can gain enough momentum in the rising edge of the intensity peak to outrun and avoid the trailing edge. Accelerating the intensity peak can further boost the momentum gain. Theory and simulations demonstrate that these dynamic intensity peaks can backwards accelerate electrons to the MeV energies required for radiation and electron diffraction probes of high energy density materials.

Vacuum laser acceleration (VLA) exploits the large electromagnetic fields of high-intensity laser pulses to accelerate electrons to relativistic energies over short distances [1-12]. The field of an intense pulse can far surpass that in conventional radio-frequency (RF) or advanced plasma-based accelerators, and the underlying interaction—involving only an electron and the electromagnetic field—has an appealing simplicity. RF accelerators routinely improve beam quality and achieve unprecedented energies, but their low damage threshold constrains the maximum accelerating field. This necessitates physically and economically immense structures to accelerate electrons to the energies necessary for high energy density probes, radiation sources such as free electron lasers, or high-energy physics experiments [13-16]. Wakefield accelerators, on the other hand, employ plasma to sustain accelerating fields nearly 1000x that of RF accelerators [17-23]. The use of plasma, however, comes with its own set of challenges, such as tuning the laser or electron beam parameters to the plasma conditions, avoiding a myriad of instabilities, and creating long uniform plasma channels [24,25].

VLA avoids damage constraints and the challenges inherent to the use of plasma, but achieving competitive electron energy gains requires a bit of ingenuity. The inherent difficulty is that the accelerating waves travel at the vacuum speed of light. As a result, electrons, regardless of their speed, will encounter repeated phases of acceleration and deceleration. More specifically, the Lawson–Woodward Theorem precludes vacuum laser acceleration under the following conditions: (1) There are no boundaries or walls present. (2) The laser-electron interaction distance and duration are infinite. (3) There are no static fields. (4) And finally, nonlinear forces, such as the magnetic Lorentz force or ponderomotive force, are ignored [1,26,27]. While in principle the Lawson–Woodward Theorem limits laser vacuum acceleration, in practice it classifies all laser vacuum acceleration schemes by which assumption(s) they exploit. In direct laser acceleration schemes, the linear electric field of a combination of laser pulses or an exotically polarized pulse accelerates injected relativistic electrons over a finite interaction length [2,7,8,11]. Crossed-beam acceleration, for example, uses the superposition of electric fields along the bisector of two linearly polarized crossing laser pulses to accelerate injected electrons [2].

Nonlinear forces become a predominant driver of electron motion in the strong electromagnetic fields characteristic of pulses delivered by modern laser systems. Accordingly, several VLA schemes utilize the ponderomotive force, which pushes electrons against the gradient of the local intensity [1,3-6,12]. For planar pulses, however, the ponderomotive force is insufficient to achieve net energy gains: The rising edge of an intensity peak that travels at the vacuum speed of light ($c$) will accelerate an electron in the direction of propagation, but the falling edge will eventually overtake and decelerate the electron back to rest [Fig. 1(a)]. To overcome this fundamental symmetry and impart net energy to an electron, the speed of the intensity peak must be subluminal, i.e. $|v_I| < c$. In vacuum beat wave acceleration, for instance, the subluminal intensity peak produced by the beating of two laser pulses with different frequencies and focusing geometries ponderomotively accelerates electrons in the direction of beat propagation [1,3].

In this Letter, we demonstrate the first vacuum acceleration of electrons in a single, planar-like laser pulse in either the forward or backward direction. This novel mechanism for VLA utilizes the "flying focus"—a recently realized spatiotemporal pulse shaping technique, in which a chirped pulse focused by a hyperchromatic diffractive optic produces an intensity peak that can propagate at any velocity, including $|v_I| < c$, over distances much longer than the Rayleigh range [28-33].

When the peak normalized vector potential of the flying focus pulse ($a_0 = eA_0/m_e c$) exceeds a critical value [$a_c = 2^{1/2}|b_I|g_I$, where $\beta_I = v_I/c$ and $g_I = (1-b_I^2)^{-1/2}$], it can accelerate electrons from rest to a final axial momentum that depends only on the velocity of the intensity peak: $p_f = 2m_e c \beta_I \gamma_I^2$. In principle, the spectral phase and power spectrum of a pulse can be adjusted to create an intensity peak with an arbitrary trajectory [28,29]. Using this principle, we also show that matching the trajectory of an intensity peak to that of an electron enhances the momentum gain beyond $2m_e c b_I g_I^2$. While this mechanism for VLA works in either the forward or backward direction, we focus on backwards acceleration due to its novelty and potential as a single pulse Compton scattering radiation source.

Figures 1(b) and 1(c) illustrate the ponderomotive acceleration of an electron in either a subluminal forward or backward flying focus intensity peak. In both cases, $a_0 > a_c$ and the electron can reach a velocity sufficient to outrun the intensity peak and retain its axial momentum, $p_f = 2m_e c \beta_I \gamma_I^2$. The laser pulse propagates from left to right at the vacuum speed of light, while the flying focus intensity peak moves independently at a velocity determined by the chirp and chromaticity of the diffractive optic (not shown). The chromatic aberration and chirp control the location and time at which each frequency comes to focus, respectively. Specifically, the intensity peak travels a distance $z_I = (\Delta w/w)f$ at a velocity $b_I = (1 \pm cT/z_I)^{-1}$, where $w$ is the central frequency of the pulse, $\Delta w/w$ its fractional bandwidth, $f$ the focal length of the diffractive optic at $w$, $T$ the stretched pulse duration, and the $\pm$ takes the sign of the chirp.

In order to demonstrate VLA in the intensity peak of a flying focus, electron dynamics, i.e.

$$\frac{d\mathbf{P}}{dt} = \frac{\partial \mathbf{a}}{\partial t} - \mathbf{v} \times (\nabla \times \mathbf{a}), \quad \text{(1a)}$$

$$\frac{d\mathbf{x}}{dt} = \boldsymbol{\beta}, \quad \text{(1b)}$$

were simulated in a model vector potential that captures the salient features of a flying focus pulse:

$$\mathbf{a} = \frac{1}{2}\hat{a}(z - \beta_I t)e^{i(z-ct)+i\eta(z-t)^2/T^2}\hat{\mathbf{x}} + \text{c.c.} \quad \text{(2)}$$

In Eqs. (1) and (2), time and space are normalized to $w$ and $k = w/c$, respectively, $\mathbf{P} = \mathbf{p}/m_e c$, $\boldsymbol{\beta} = \mathbf{P}/\gamma$, $g = (1+P^2)^{1/2}$, $\hat{\mathbf{x}}$ is the unit vector in the x-direction, $h = [(aT\Delta w)^2 - 1]^{1/2}$ is the chirp parameter, and $a$ depends on the power spectrum (e.g. $a = [8\ln(2)]^{-1/2}$ for a Gaussian). The amplitude, $\hat{a}(z - \beta_I t)$, represents the intensity peak traveling at the velocity, $v_I$. The shape of the peak was taken as a fifth order polynomial so that its value and derivative vanish at the leading and trailing edges. While Eq. (2) includes the phase contributed by the chirp, it was not found to affect the results. In each simulation, electrons were initialized at rest.

Figure 2 displays the results of these simulations and illustrates the principal features of ponderomotive acceleration in a backwards travelling, subluminal intensity peak. Below the cutoff vector potential determined by the speed of the intensity peak, $a_0 < a_c = 2^{1/2} |\beta_I| \gamma_I$, an electron gains insufficient axial momentum to outrun the peak. The peak overtakes the electron and the electron loses all of its axial momentum. Above the cutoff vector potential ($a_0 > a_c$), on the other hand, an electron gains enough momentum to outrun the peak and retains its energy. In this case, the final momentum depends only on the velocity of the intensity peak, i.e. it is independent of the vector potential ($P_z = 2 b_I g_I^2$). The final momentum of an electron increases with the velocity of the intensity peak and even diverges as $b_I \to 1$. However, the required vector potential increases as well. Operating at the lowest possible vector potential ($a_0 = a_c$) provides the scaling $|P_z| = a_0 (2 + a_0^2)^{1/2}$.

The underlying mechanism behind the acceleration can be understood as a reflection from the ponderomotive potential of the flying focus intensity peak in its Lorentz frame. Equations (1) and (2) admit two lab frame conservation relations. The first equates the transverse momentum to the vector potential, $P_x = a_x$, and the second the phase-averaged axial momentum to the energy:

$$\frac{d}{dt}(\langle\gamma\rangle - \beta_I \langle P_z \rangle) = 0, (3)$$

where $\langle\rangle$ denotes an average over the rapidly varying phase of the vector potential and $\langle\gamma\rangle = [1 + \frac{1}{2}\hat{a}^2 + \langle P_z \rangle^2]^{1/2}$. Multiplying Eq. (3) by $g_I$ and recognizing the result as the electron energy in the frame of the intensity peak, i.e. $\langle g' \rangle = g_I \langle g \rangle - b_I g_I \langle P_z \rangle$, gives

$$\frac{d\langle\gamma'\rangle}{dt} = 0. (4)$$

That is the electron energy in the peak frame is conserved.

In this frame, the electron dynamics reduce to an interaction with a stationary ponderomotive potential barrier (Fig. 3). Only two outcomes satisfy Eq. (4) after the electron-barrier interaction: (1) The initial kinetic energy of the electron is insufficient to overcome the potential barrier and the electron is reflected [Fig. 3(a)], or (2) the initial kinetic energy is sufficient to overcome the potential barrier [Fig. 3(b)]. Upon Lorentz transforming back into the lab frame, the first case results in net energy gain as the electron overtakes the intensity peak [Fig. 3(c)], while the second case results in zero net energy gain as the intensity peak overtakes the electron [Fig. 3(d)].

The cutoff vector potential required for a reflection, and hence energy gain, can be found by recasting Eq. (4) in terms of the axial momentum. For an electron initially positioned outside of the laser pulse, the phase average axial momentum in the peak frame is

$$\langle P'_z \rangle = \pm\sqrt{(\beta_I - \beta_0)^2 \gamma_0^2 \gamma_I^2 - \tfrac{1}{2}\hat{a}^2} \quad (5)$$

where $b_0$ is the initial velocity of the electron in the lab frame, and $g_0 = (1 - b_0^2)^{-1/2}$. The positive and negative roots in Eq. (5) correspond to the electron overtaking the potential barrier and reflecting from the barrier, respectively [Figs. 3(a) and (b)]. At the point of reflection, i.e. the turning point, the axial momentum must vanish. As a result, the cutoff value of $a_0$ needed to reflect the electron satisfies Eq. (5) with $\langle P'_z \rangle = 0$:

$$a_c = 2^{1/2} \, | \beta_I - \beta_0 | \, \gamma_0 \gamma_I. \quad (6)$$

For an electron initially at rest, this simplifies to $a_c = 2^{1/2} \beta_I \gamma_I$. When $a_0 > a_c$, the electron retains a lab frame momentum $\langle P_z \rangle = [2\beta_I - \beta_0(1+\beta_I^2)]\gamma_0 \gamma_I^2$ upon exiting the pulse. If $g_0 = 1$, $\langle P_z \rangle = 2\beta_I \gamma_I^2$. Plots of this final momentum as a function of $b_I$ and conditional on $a_0 > a_c$ are indistinguishable from Fig. (2).

The derivation of Eq. (3) required averaging over the rapid spatiotemporal oscillations in the electron motion caused by the phase of the vector potential. The validity of this averaging holds when electrons quickly pass through many optical cycles, allowing for a clean separation of the time scales associated with intensity and phase variations. While this is always the case for electrons moving in the opposite direction of the phase velocity, highly relativistic electrons moving in the same direction as the phase velocity can experience a near-constant phase of the vector potential for extended durations and distances, which can blur the timescales of phase and intensity variations. A rough validity condition for the averaging can be found by ensuring that an electron experiences many doppler-shifted cycles during the interaction, i.e. $w(1 - b_z)T_I \gg 2p$, where $T_I$ is the interaction time. For $a_0 \gg 1$, this simplifies to $wT_I \gg 4pa_0^4$, while for $a_0 \ll 1$, $wT_I \gg 2p$. Note that this is an important distinction between vacuum ponderomotive acceleration and ponderomotive acceleration in a plasma (c.f. Ref. [34]). In a sufficiently dense plasma, the superluminal phase velocity of the light wave justifies the use of Eq. (3) for codirectional, highly relativistic electrons regardless of the peak vector potential.

Accelerating the intensity peak to match the ponderomotive acceleration of an electron, i.e. "trajectory locking," can substantially increase the momentum gain. The constant velocity intensity peaks considered above limited the interaction distance and the momentum gain to a value determined by the maximum vector potential (Fig. 2). By limiting the interaction distance, the constant velocity scheme wastes any length that the intensity peak has yet to travel. Trajectory locking, on the other hand, keeps the electron in the ponderomotive potential and can utilize the entire distance, $z_I$, to increase the final momentum (Fig. 4).

In the trajectory-locked (TL) scheme, the intensity peak initially moves at a constant velocity, $b_{I0}$. Once the electron has accelerated from rest to the velocity of the intensity peak, which occurs at the location $a = a_c$, the intensity peak accelerates to keep the electron at this location (i.e. at $a = a_c$) (Fig. 4). The momentum of the electron evolves according to the ponderomotive guiding center [35] equation of motion,

$$\frac{d\langle P_z \rangle}{dt} = -\frac{1}{4\langle \gamma \rangle}\frac{\partial}{\partial z}a^2(z,t). \quad (7)$$

Because the electron samples a constant intensity gradient, Eq. (7) can be directly integrated to find the momentum,

$$\langle P_z(t > t_c)\rangle \approx \sqrt{(\beta_{I0}\gamma_{I0}^2)^2 + \frac{1}{2}(t - t_c)\frac{\partial a^2}{\partial z}\bigg|_{a=a_c}}, \quad (8)$$

where $\beta_{I0}\gamma_{I0}^2$ is the phase average electron momentum in the lab frame upon reaching $a_c$ at time $t_c$. As expected, Eq. (8) predicts that optimizing the momentum gain requires co-locating $a_c$ with the maximum intensity-gradient of the peak. Asymptotically, the momentum gain has a relatively weak scaling with time, $\langle P_z \rangle \propto t^{1/2}$. This results from the diminished ponderomotive force as $\langle g \rangle$ increases [RHS of Eq. (7)]. Rewriting Eq. (8) provides the electron velocity and, correspondingly, the peak velocity needed to accomplish trajectory-locking:

$$\langle \beta_z(t > t_c) \rangle \approx \frac{\langle P_z(t > t_c) \rangle}{\sqrt{1 + \frac{1}{2} a_c^2 + \langle P_z(t > t_c) \rangle^2}}. \quad (9)$$

To demonstrate electron acceleration in a TL peak, electron dynamics were simulated using Eq. (7) and the z-component of Eq. (1b). Figure 5(a) compares an example of the momentum gain in a TL peak to that in a constant velocity peak. At the breakeven time, $T_{BE}$, the momentum gain in the TL scheme equals the asymptotic momentum gain, $2\beta_{I0}\gamma_{I0}^2$, in the constant velocity scheme. At time $T_E$, an electron accelerated by the constant velocity peak has exited the peak and reached its asymptotic momentum gain, while an electron in the TL peak continues to gain momentum. As shown in Fig. 5(b), an electron in a TL peak reaches $\langle P_z \rangle = 2\beta_{I0}\gamma_{I0}^2$ in a fraction of the time it would take in a constant velocity peak over a wide range of parameters. Alternatively, Fig. 5(c) shows that the momentum gain in the TL scheme can exceed that in the constant velocity scheme over a time $T_E$. The exception occurs when $a_0 \gg a_c$; near the maximum vector potential, the intensity gradient becomes too small for effective TL acceleration. Note that Fig. 5(c) represents the momentum gain after a time $T_E$ and not the maximum achievable momentum gain in the TL scheme, which is only limited by the interaction length. Figures 5(b) and (c) are independent of the duration of the intensity peak.

The acceleration schemes described here considered electrons located on the propagation axis at the center of the laser spot. These electrons experience zero transverse ponderomotive force and remain on the axis indefinitely. Electrons initially located off-axis would undergo lateral motion in response to the transverse ponderomotive force. In the case of a transverse Gaussian profile (or any profile that decreases with radius), the ponderomotive force would radially expel electrons well before significant longitudinal acceleration could occur. However, this can be avoided by using two polarizations and multiple spatial modes to shape the transverse profile of the intensity peak. For instance, a superposition of Hermite-Gaussian modes with appropriate polarizations, as described in Refs. [4,6,36], can create a radial ponderomotive potential well that focuses off axis electrons to the center of the laser spot, preventing radial expulsion. Because these modes are eigenfunctions of the paraxial wave equation, the intensity profile will retain its shape throughout propagation.

A novel concept for VLA based on the recently demonstrated "flying focus" offers a tunable source of high-energy electrons that can travel in either the forward or backward direction. Both the energy and direction can be controlled by adjusting the chirp of a laser pulse focused by a diffractive optic. The resulting intensity peak, and hence ponderomotive force, can travel at any velocity over distances unconstrained by diffraction. A subluminal intensity peak, traveling either forwards or backwards, breaks a fundamental symmetry in the interaction of an electron with a planar-like laser pulse, allowing for net energy gain: Electrons ponderomotively accelerated by a subluminal intensity peak can outrun the peak and retain their momentum. In the Lorentz frame of the intensity peak, this is simply a reflection from the ponderomotive potential. By adjusting the spectral phase of the laser pulse, the intensity peak can be accelerated and matched to the trajectory of the electron, providing enhanced energy gains. The case of backwards acceleration described here enables a promising scheme for all-optical, single pulse inverse Compton scattering [37-39]. The intensity peak accelerates electrons against the phase velocity, which will result in the same pulse Compton scattering off the electrons. This avoids the issues of alignment and independently preparing high-energy electrons and a counter-propagating pulse. As an example, a 10 MeV electron will radiate light with an upshifted frequency nearly 1500 times that of the incident light.

**Acknowledgements**

The authors would like to thank J.L. Shaw, R.K. Follett, and D. Turnbull for fruitful discussions.

The work published here was supported by the US Department of Energy Office of Fusion Energy Sciences under contract no. DE-SC0016253, the Department of Energy under cooperative agreement no. DE-NA0003856, the University of Rochester, and the New York State Energy Research and Development Authority.

or represents that its use would not infringe privately owned rights. Reference herein to any specific commercial product, process, or service by trade name, trademark, manufacturer, or otherwise does not necessarily constitute or imply its endorsement, recommendation, or favoring by the U.S. Government or any agency thereof.

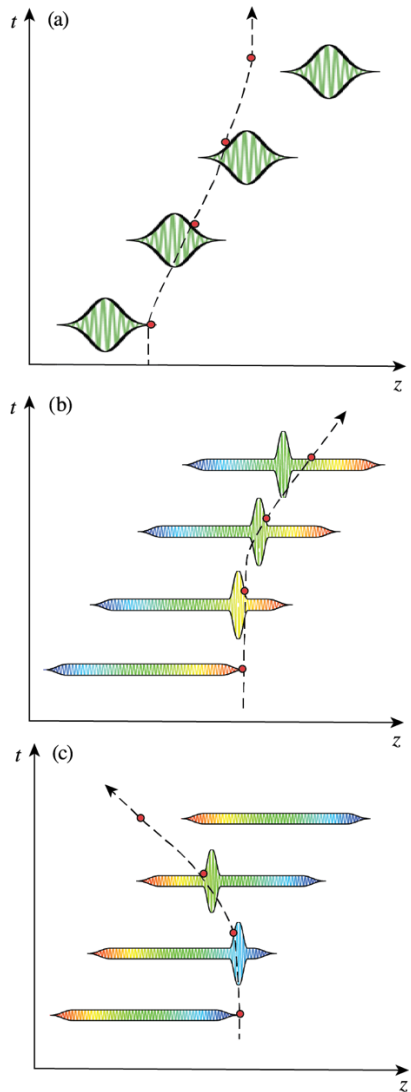

FIG. 1. (a) A typical luminal intensity peak in vacuum. The electron, shown as a red dot, experiences equal and opposite ponderomotive accelerations on the leading and falling edges of the pulse respectively and gains no net energy. (b) A positively chirped flying focus with a subluminal intensity peak. After forward acceleration in the leading edge of the intensity peak, the electron outruns the peak and retains the energy it gained. (c) A negatively chirped flying focus with a subluminal intensity peak that travels in the opposite direction of the pulse. After backwards acceleration in the leading edge, the electron outruns the intensity peak and retains the energy it gained.

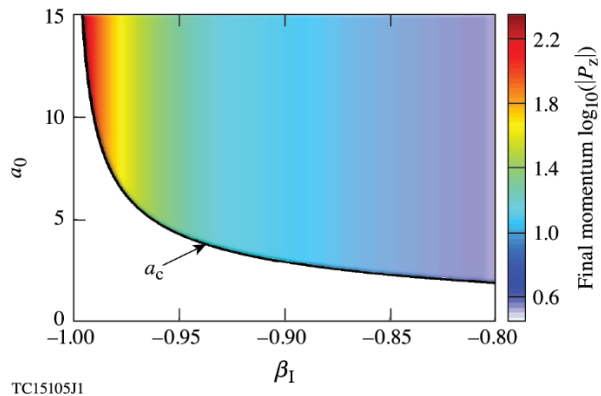

TC15105J1

FIG. 2. Final momentum of an electron accelerated in a backwards propagating flying focus intensity peak. Below the cutoff vector potential ($a_c$), an electron acquires a velocity insufficient to outrun the intensity peak. Above the cutoff, an accelerated electron can outrun the intensity peak, and the final momentum is independent of $a_0$.

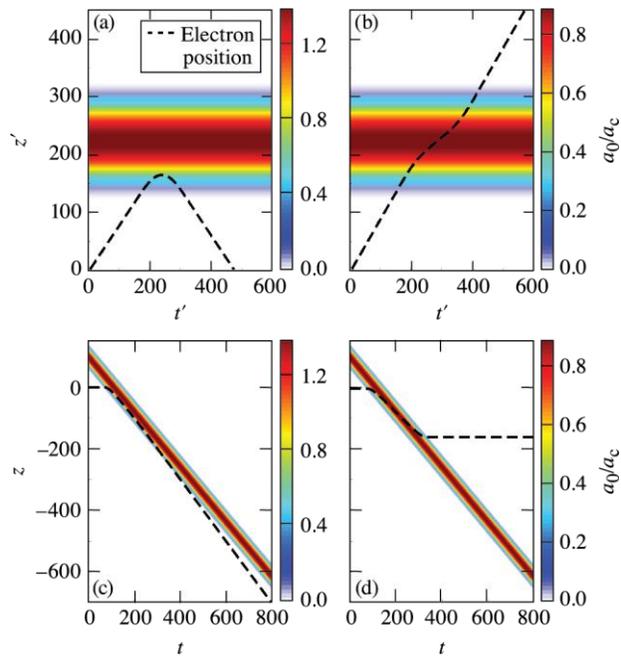

FIG. 3. Trajectories of an electron (black dashed lines) and intensity peak (contours) in the Lorentz frame of the intensity peak, (a) and (b), and the lab frame, (c) and (d). In (a) the normalized vector potential exceeds the cutoff, and the ponderomotive potential reflects the electron, resulting in net momentum gain in the lab frame (c). In (b) the ponderomotive potential is too weak to reflect the electron and it returns to rest in the lab frame (d).

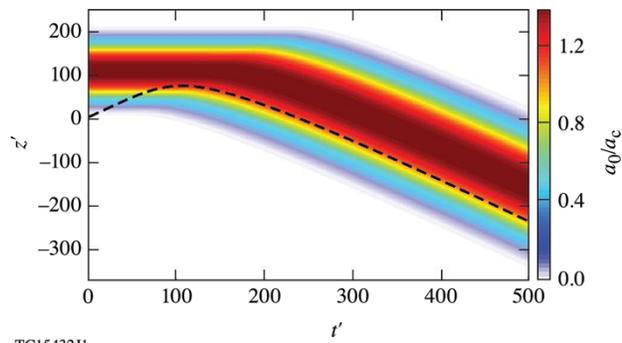

FIG. 4. Path of an electron (black dashed line) and a trajectory-locked intensity peak (contours) in the Lorentz frame associated with the initial velocity of the intensity peak. The intensity peak is accelerated to keep the electron at the initial cutoff vector potential as its momentum increases.

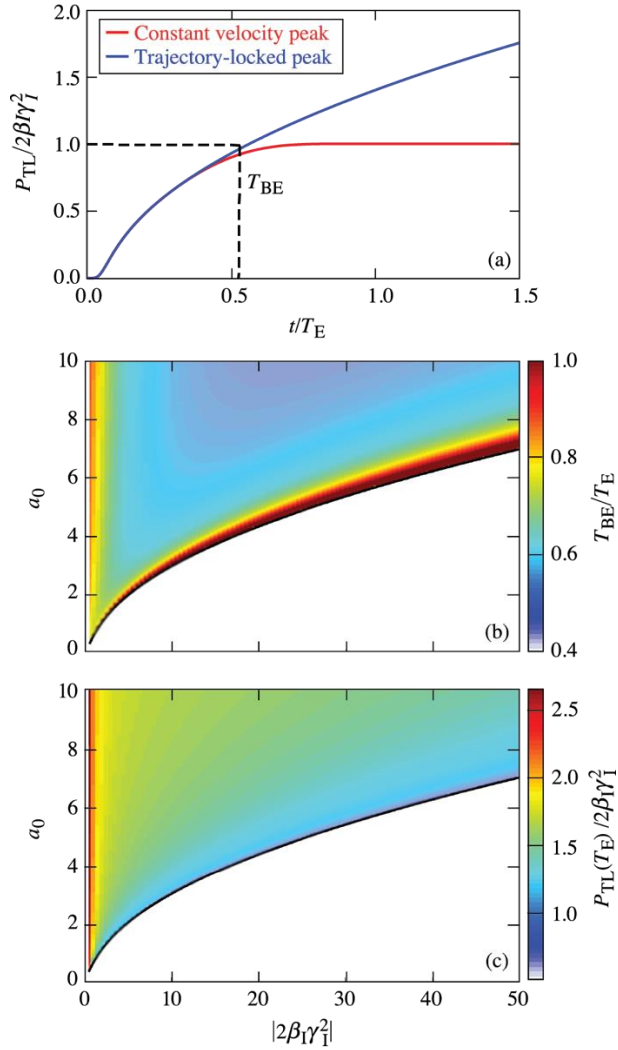

TC15433J1

FIG. 5. (a) Comparison of the momentum gained by an electron in trajectory-locked and constant velocity intensity peaks. An electron exits the constant velocity peak and reaches its asymptotic momentum, $2b_{I0}g_{I0}^2$, at time $T_E$. In the trajectory-locked peak, the electron reaches a momentum $2b_{I0}g_{I0}^2$ at the breakeven time $T_{BE}$ and continues to accelerate. (b) An electron in a trajectory-locked peak reaches a momentum $2b_{I0}g_{I0}^2$ in a fraction of the time it would take in the constant velocity peak over a wide range of parameters. (c) At the constant velocity exit time, an electron in a trajectory-locked peak typically reaches a higher momentum than one in a constant velocity peak.